\documentclass[fleqn,twoside]{article}
\usepackage{espcrc2}

\usepackage{graphicx}
\usepackage[figuresright]{rotating}

\title{Light Scalar Mesons from $\phi$ to $\psi$}

\author{N.N. Achasov \address[MCSD]{Laboratory of Theoretical Physics,
S.L. Sobolev Institute for Mathematics, 630090, Novosibirsk, Russian
Federation}, A.V. Kiselev \addressmark{\tt}, and G.N. Shestakov
\addressmark{\tt}}

\begin{document}

\begin{abstract}
The following topics are considered.
\begin{enumerate}
\item Confinement, chiral dynamics, and light scalar
mesons
\item Chiral shielding of the $\sigma(600)$
\item The $\phi$ meson radiative decays about nature of light scalar
resonances
\item The $J/\psi$ decays about nature of light scalar
resonances
\item The $a_0(980)\to\gamma\gamma$ and
$f_0(980)\to\gamma\gamma$ decays about nature of light scalar
resonances
\item New round in $\gamma\gamma\to\pi^+\pi^-$, the Belle data
\end{enumerate}

Arguments in favor of the four-quark model of the $a_0(980)$ and
$f_0(980)$ mesons are given.

\vspace{1pc}
\end{abstract}

\maketitle

\section{INTRODUCTION. CONFINEMENT, CHIRAL DYNAMICS, AND LIGHT SCALAR MESONS}

The scalar channels in the region up to 1 GeV became a stumbling
block of QCD. The point is that both perturbation theory and sum
rules do not work in these channels because there are not solitary
resonances in this region. At the same time the question on the
nature of the light scalar mesons is major for  understanding the
mechanism of the chiral symmetry realization, arising from the
confinement,  and hence for  understanding the confinement itself.
In the talk are discussed the chiral shielding of the $\sigma(600)$,
$\kappa(800)$ mesons, a role of the radiative $\phi$ decays, the
heavy quarkonia decays, the $\gamma\gamma$ collisions in decoding
the nature of the light scalar mesons and evidence in favor of the
four-quark nature of the light scalar mesons. New goal and
objectives are considered also.

To discuss actually the nature of the nonet of the light scalar
mesons: the  putative $f_0(600)$ (or $\sigma (600)$) and $\kappa
(700-900)$ mesons and the well-established $f_0(980)$ and $a_0(980)$
mesons, one should explain not only their mass spectrum,
particularly the mass degeneracy of the $f_0(980)$ and $a_0(980)$
states, but answer the next real challenges.
\begin{enumerate}
\item The copious $\phi\to\gamma f_0(980)$ decay
and especially the copious $\phi\to\gamma a_0(980)$ decay, which
looks as the decay plainly forbidden by the Okubo-Zweig-Iizuka (OZI)
rule in the quark-antiquark model of $a_0(980)=(u\bar u - d\bar
d)/\sqrt{2}$.
\item Absence of $J/\psi\to a_0(980)\rho$ and
$J/\psi\to f_0(980)\omega$ with copious $J/\psi\to a_2(1320)\rho$,
$J/\psi\to f_2(1270)\omega$ if $a_0(980)$ and $f_0(980)$ are $P$
wave states of $q\bar q$ like $a_2(1320)$ and $f_2(1270)$
respectively.
\item Absence of $J/\psi\to\gamma f_0(980)$ with copious
$J/\psi\to\gamma f_2(1270)$ and $J/\psi\to\gamma f_2^\prime
(1525)\phi$ if $f_0(980)$ is $P$ wave state of $q\bar q$ like
$f_2(1270)$ or $f_2^\prime(1525)$.
\item Suppression of $a_0(980)\to\gamma\gamma$ and $f_0(980)\to
\gamma\gamma$ with copious $a_2(1320)\to\gamma\gamma$,
$f_2(1270)\to\gamma\gamma$ if $a_0(980)$ and $f_0(980)$ are $P$ wave
state of $q\bar q$ like $a_2(1320)$and $f_2(1270)$ respectively.
\end{enumerate}

As already noted, the study of the nature of light scalar resonances
has become a central problem of non-perturbative QCD because the
elucidation of their nature is important for understanding the
chiral symmetry realization way in the low energy region, i.e., the
main consequence of QCD in the hadron world. As Experiment suggests,
Confinement forms colourless observable hadronic fields and
spontaneous breaking of chiral symmetry with massless pseudoscalar
fields. There are two possible scenarios for QCD at low
energy.\vspace*{0.15cm}

1. Non-linear $\sigma$-model,\\
$L=\left(F_\pi^2/2\right)Tr\left(\partial_\mu V(x)\partial^\mu
V^+(x)\right) +...,$ where\\ $ V(x)=\exp \{2\imath\phi(x)/F_\pi\}.$
\vspace*{0.15cm}

2. Linear $\sigma$-model,\\
$L=\frac{1}{2}Tr\left(\partial_\mu \mbox{V}(x)\partial^\mu
\mbox{V}^+(x)\right)- W\left(\mbox{V}(x)\mbox{V}^+(x)\right),$\\
where $\mbox{V}(x)=(\sigma(x)+\imath\pi(x)).$ \vspace*{0.15cm}

The experimental nonet of the light scalar mesons, the putative
$f_0(600)$ (or $\sigma (600)$) and $\kappa (700-900)$ mesons and the
well-established $f_0(980)$ and $a_0(980)$ mesons as if suggests the
$U_L(3)\times U_R(3)$ linear $\sigma$ model. Hunting the light
$\sigma$ and $\kappa$ mesons had begun in the sixties already and a
preliminary information on the light scalar mesons in Particle Data
Group (PDG) Reviews had appeared at that time. But long-standing
unsuccessful attempts to prove their existence in a conclusive way
entailed general disappointment and an information on these states
disappeared from PDG Reviews. One of principal reasons against the
$\sigma$ and $\kappa$ mesons was the fact that both $\pi\pi$ and
$\pi K$ scattering phase shifts do not pass over $90^0$ at putative
resonance masses.

\section{CHIRAL SHIELDING OF THE \boldmath{$\sigma(600)$} \cite{AS1,ADSK,AK}}

Situation changes when we showed that in the linear $\sigma$
model \cite{GML},\\
$L=\left(1/2\right)\left
[(\partial_\mu\sigma)^2+(\partial_\mu\overrightarrow{\pi})^2\right]
+(\mu^2/2)[(\sigma)^2+$\\
$(\overrightarrow{\pi})^2]-(\lambda/4)
[(\sigma)^2+(\overrightarrow{\pi})^2]^2,$\\ there is a negative
background phase which hides the $\sigma$ meson \cite{AS1}.
\begin{figure}\includegraphics[width=17pc]{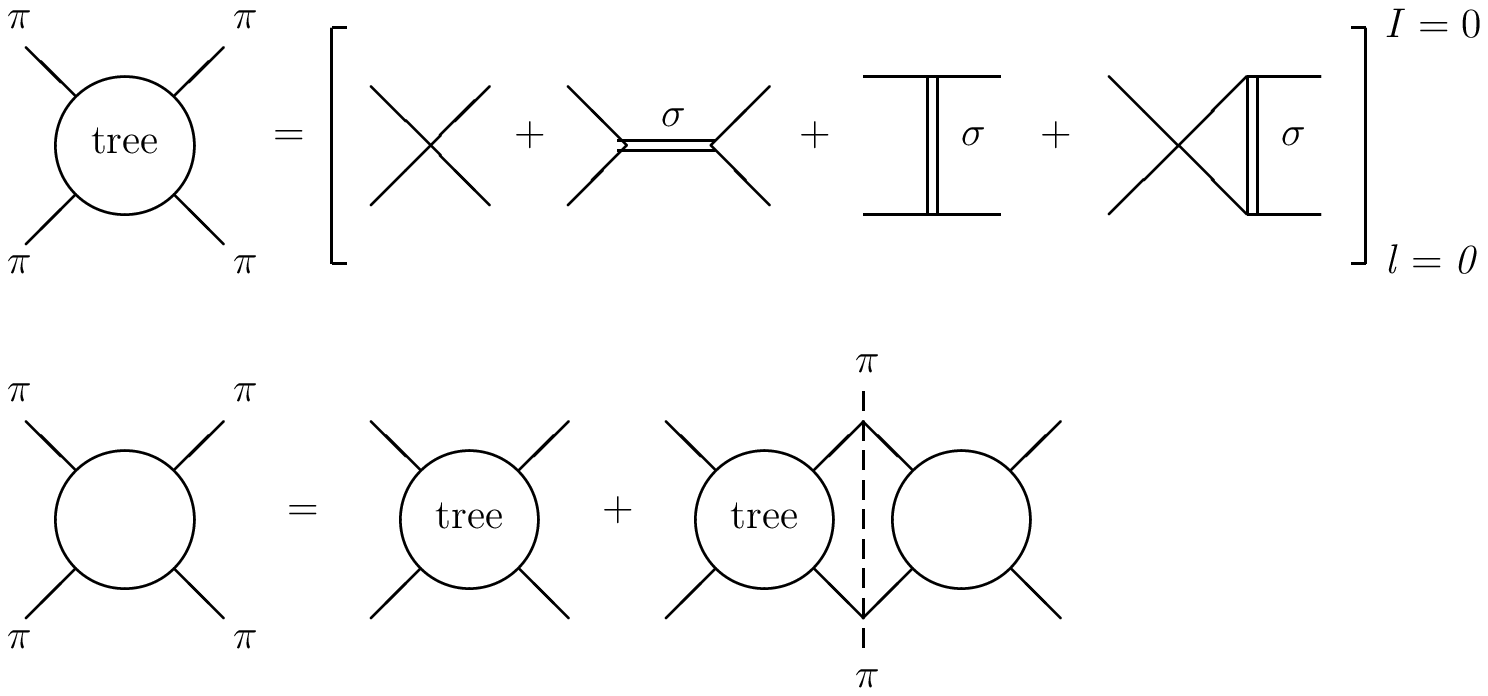}
\caption{\small The graphical representation of the $S$ wave $I=0$
$\pi\pi$ scattering amplitude.}
\end{figure}
It has been made clear that shielding wide lightest scalar mesons in
chiral dynamics is very natural. This idea was picked up and
triggered new wave of theoretical and experimental searches for the
$\sigma$ and $\kappa$ mesons. According the simplest Dyson equation
for the $\pi\pi$ scattering amplitude with real intermediate
$\pi\pi$ states only, see Fig. 1,
\begin{eqnarray}T^0_0=\frac{T_0^{0(tree)}}{1-\imath\rho_{\pi\pi}T_0^{0(tree)}}=
\frac{e^{2\imath\delta^0_0}-1}{2\imath
\rho_{\pi\pi}}\nonumber\end{eqnarray}
$$=\frac{1}{\rho_{\pi\pi}}\left[\left
(\frac{e^{2\imath\delta_{bg}}-1}{2\imath}\right)+
e^{2\imath\delta_{bg}}T_{res}\right]\,,$$
\begin{eqnarray}
T_{res}=\frac{\sqrt{s}\Gamma_{res}(s)}{M^2_{res} - s +
\Re(\Pi_{res}(M^2_{res}))- \Pi_{res}(s)}\nonumber\end{eqnarray}
$$=(e^{2\imath\delta_{res}}-1)/(2\imath\rho_{\pi\pi})\,,$$
$$T_{bg}=\frac{\lambda(s)}{1-\imath\rho_{\pi\pi}\lambda(s)}
=\frac{e^{2\imath\delta_{bg}}-1}{2\imath
\rho_{\pi\pi}} \ ,\quad\lambda(s)= $$
$$\frac{m_\pi^2-m_\sigma^2}{32\pi
F^2_\pi}\left [5-2\frac{m_\sigma^2-m_\pi^2}{s-4m_\pi^2}\ln\left
(1+\frac{s-4m^2_\pi}{m_\sigma^2}\right )\right ]\,,\
$$ $$\delta^0_0=\delta_{bg}+\delta_{res}\,,$$
$$\Im(\Pi_{res}(s))=\sqrt{s}\Gamma_{res}(s)=\frac{3}{2}\frac{g_{res}^2(s)}
{16\pi}\rho_{\pi\pi}\,,$$
$$ \Re(\Pi_{res}(s))=-3\frac{g_{res}^2(s)}{32\pi}\lambda(s)
\rho_{\pi\pi}^2\,,$$ $$
g_{res}(s)=\frac{g_{\sigma\pi^+\pi^-}}{|1-\imath\rho_{\pi\pi}\lambda(s)|}\
,$$ $$M^2_{res}= m_\sigma^2 - \Re(\Pi_{res}(M^2_{res})),\
\rho_{\pi\pi}=\sqrt{1-4m_\pi^2/s},$$ where $s=m^2$ and $m$ is the
invariant mass of the $\pi\pi$ system. These simple formulae show
that the resonance contribution is strongly modified by the chiral
background amplitude.

In theory the principal problem is impossibility to use the linear
$\sigma$ model in the tree level approximation inserting widths into
$\sigma$ meson propagators because such an approach breaks the both
unitarity and Adler self-consistency conditions. Strictly speaking,
the comparison with the experiment requires the non-perturbative
calculation of the process amplitudes. Nevertheless, now there are
the possibilities to estimate odds of the $U_L(3)\times U_R(3)$
linear $\sigma$ model to underlie physics of light scalar mesons in
phenomenology. Really, even now there is a huge body of information
about the $S$ waves of different two-particle pseudoscalar states.
As for theory, we know quite a lot about the scenario under
discussion: the nine scalar mesons, the putative chiral shielding of
the $\sigma (600)$ and $\kappa(700-900)$ mesons, the unitarity,
analiticity and Adler self-consistency conditions. In addition,
there is the light scalar meson treatment motivated by field theory.
The foundations of this approach were formulated in our papers
\cite{ADSK}. In particular, in this approach  were introduced
propagators of scalar mesons, satisfying the K\"allen-Lehmann
representation. Recently \cite{AK}, the comprehensive examination of
the chiral shielding of the $\sigma (600)$ has been performed with a
simultaneous analysis of the modern data on the
$\phi\to\gamma\pi^0\pi^0$ decay and the classical $\pi\pi$
scattering data. Figures 2, 3, and 4 show an example of the fit to
the data on the $S$ wave $I=0$ $\pi\pi$ scattering phase shift
$\delta_0^0=\delta_B^{\pi\pi}+ \delta_{res}$ and the resonance
($\delta_{res}$) and background ($\delta_B^{\pi\pi}$) components of
$\delta_0^0$, respectively (all the phases in degrees).
\begin{figure}\includegraphics[width=17pc]{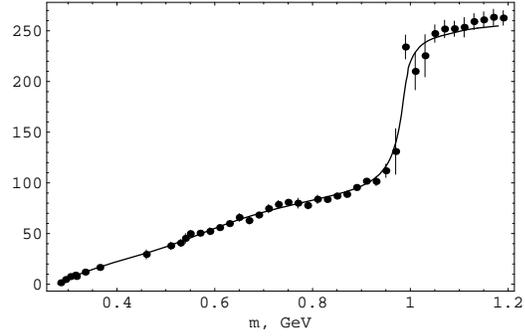}
\caption{\small The $S$ wave $I=0$ $\pi\pi$ scattering phase shift
$\delta_0^0$.}\end{figure}
\begin{figure}\includegraphics[width=17pc]{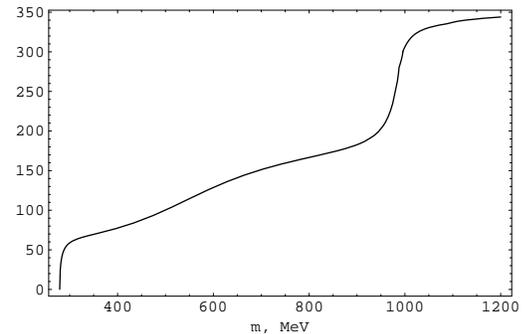}
\caption{\small The resonance phase shift
$\delta_{res}$.}\end{figure}
\begin{figure}\includegraphics[width=17pc]{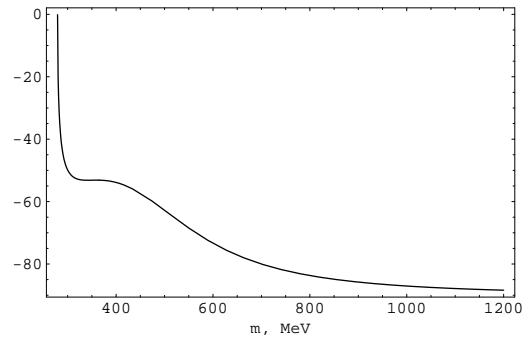}
\caption{\small The background phase shift $\delta_B^{\pi\pi}$}
\end{figure}
An example of the fit to the $\phi\to\gamma\pi^0\pi^0$ data in this
case is shown in Fig. 6.

\section{FOUR-QUARK MODEL}

The nontrivial nature of the well-established light scalar
resonances $f_0(980)$ and $a_0(980)$ is no longer denied practically
anybody. In particular, there exist numerous evidences in favour of
the $q^2\bar q^2$ structure of these states \cite{A1,A2}. As for the
nonet as a whole, even a look at PDG Review gives an idea of the
four-quark structure of the light scalar meson nonet, $\sigma
(600)$, $\kappa (700-900)$, $f_0(980)$, and $a_0(980)$, inverted in
comparison with the classical $P$ wave $q\bar q$ tensor meson nonet,
$f_2(1270)$, $a_2(1320)$, $K_2^\ast(1420)$, $f_2^\prime (1525)$.
\footnote{To be on the safe side, notice that the linear $\sigma$
model does not contradict to non-$q\bar q$ nature of the low lying
scalars because Quantum Fields can contain different virtual
particles in different regions of virtuality.}

Really, while the scalar nonet cannot be treated as the $P$ wave
{$q\bar q$ nonet in the naive quark model, it can be easy understood
as the $q^2\bar q^2$ nonet, where $\sigma (600)$ has no strange
quarks, $\kappa (700-900)$ has the $s$ quark, $f_0(980)$ and
$a_0(980)$ have the $s\bar s$ pair \cite{J,BFSS}. The scalar mesons
$a_0(980)$ and $f_0(980)$, discovered more than thirty years ago,
became the hard problem for the naive $q\bar q$ model from the
outset. Really, on the one hand the almost exact degeneration of the
masses of the isovector $a_0(980)$ and isoscalar $f_0(980)$ states
revealed seemingly the structure similar to the structure of the
vector $\rho$ and $\omega$ or  tensor $a_2(1320)$ and $f_2(1270)$
mesons, but on the other hand the strong coupling of $f_0(980)$ with
the $K\bar K$ channel as if suggested a considerable part of the
strange pair $s\bar s$ in the wave function of $f_0(980)$. In 1977
R.L. Jaffe \cite{J} noted that in the MIT bag model, which
incorporates confinement phenomenologically, there are light
four-quark scalar states. He suggested also that $a_0(980)$ and
$f_0(980)$ might be these states with symbolic structures:
$a^0_0(980)=(us\bar u\bar s-ds\bar d\bar s)/\sqrt{2}$, and
$f_0(980)=(us\bar u\bar s + ds\bar d\bar s)/\sqrt{2}$. From that
time $a_0(980)$ and $f_0(980)$ resonances came into beloved children
of the light quark spectroscopy.

\section{RADIATIVE DECAYS OF \boldmath{$\phi$} MESON ABOUT NATURE OF LIGHT
SCALAR RESONANCES \cite{AI,A2,AG}}

Ten years later we showed \cite{AI} that the study of the radiative
decays $\phi\to\gamma a_0\to\gamma\pi\eta$ and $\phi\to\gamma f_0\to
\gamma\pi\pi$ can shed light on the problem of $a_0(980)$ and
$f_0(980)$ mesons. Over the next ten years before experiments (1998)
the question was considered from different points of view. Now these
decays have been studied not only theoretically but also
experimentally. The first measurements have been reported by the SND
and CMD-2 Collaborations which obtain the following branching
ratios\\ $\ \ B(\phi\to\gamma\pi^0\eta)= (0.88\pm 0.14\pm
0.09)\cdot10^{-4}\,,$\\
$\ \ B(\phi\to\gamma\pi^0\pi^0)= (1.221\pm 0.098\pm
0.061)\cdot10^{-4}\,,$\\
$\ \ B(\phi\to\gamma\pi^0\eta)=(0.9\pm 0.24\pm 0.1)\cdot10^{-4}\,,$\\
$\ \ B(\phi\to\gamma\pi^0\pi^0)=(0.92\pm0.08\pm0.06)\cdot10^{-4}\,.$\\
More recently the  KLOE Collaboration has measured\\
$\ \ B(\phi\to\gamma\pi^0\eta)=(0.851\pm0.051\pm0.057)\cdot
10^{-4}\,,$\\
$\ \ B(\phi\to\gamma\pi^0\eta
)=(0.796\pm0.060\pm0.040)\cdot10^{-4}\,,$\\
$\ \ B(\phi\to\gamma\pi^0\pi^0)=
(1.09\pm0.03\pm0.05)\cdot10^{-4}$\\
in agreement with the Novosibirsk data but with a considerably
smaller error. Note that $a_0(980)$  is produced in the radiative
$\phi$ meson decay as intensively as $\eta '(958)$  containing
$\approx 66\% $ of $s\bar s$, responsible for $\phi\approx s\bar
s\to\gamma s\bar s\to\gamma \eta '(958)$. It is a clear qualitative
argument for the presence of the $s\bar s$ pair in the isovector
$a_0(980)$ state, i.e., for its four-quark nature.

When basing the experimental investigations, we suggested one-loop
model $\phi\to K^+K^-\to\gamma a_0(980)$ (or $f_0(980))$ \cite{AI}.
This model is used in the data treatment and is ratified by
experiment. Below we argue on gauge invariance grounds that the
present data give the conclusive arguments in favor of the $K^+K^-$
loop transition as the principal mechanism of $a_0(980)$ and
$f_0(980)$ meson production in the $\phi$ radiative decays
\cite{A2,AG}. This enables to conclude that production of the
lightest scalar mesons $a_0(980)$ and $f_0(980)$ in these decays is
caused by the four-quark transitions, resulting in strong
restrictions on the large $N_C$ expansions of the decay amplitudes.
The analysis shows that these constraints give new evidences in
favor of the four-quark nature of $a_0(980)$ and $f_0(980)$ mesons
\cite{A2}. The data are described in the model $\phi\to(\gamma
a_0+\pi^0\rho)\to\gamma\pi^0\eta$ and $\phi\to [\gamma
(f_0+\sigma)+\pi^0\rho]\to\gamma\pi^0\pi^0$. The resulting fits to
the KLOE data are presented in Figs. 5 and 6.

\begin{figure}\includegraphics[width=17pc]{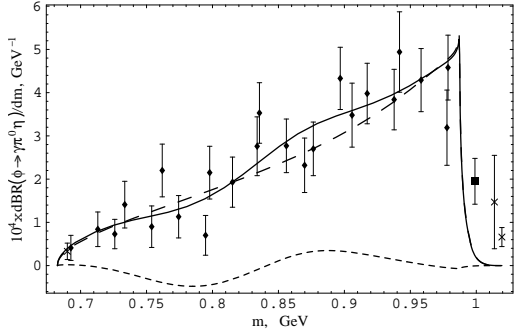}
\caption{\small The fit to the KLOE data for the $\pi^0\eta$ mss
spectrum in the $\phi\to\gamma\pi^0\eta$ decay.}\end{figure}

To describe the experimental spectra $$S_R(m)\equiv dB(\phi\to\gamma
R\to\gamma ab\,,\, m)/dm$$
\begin{eqnarray}&& =\frac{2m^2\Gamma(\phi\to\gamma
R\,,\,m)\Gamma(R\to ab\,,\,m)}{\pi\Gamma_\phi|D_R(m)|^2}\nonumber\\
\hspace*{-9mm}&& = \frac{4|g_R(m)|^2\omega (m)
p_{ab}(m)}{\Gamma_\phi\, 3(4\pi)^3m_{\phi}^2}\left
|\frac{g_{Rab}}{D_R(m)}\right |^2\nonumber
\end{eqnarray} (where $1/D_R(m)$ and  $g_{Rab}$ are the propagator
and coupling constants of $R=a_0,f_0$; $ab=\pi^0\eta$, $\pi^0\pi^0$)
the function $|g_R(m)|^2$ should be smooth, almost constant, in the
range $m\leq 0.99$ GeV. But the problem issues from gauge invariance
which requires that $A [\phi(p)\to\gamma (k) R(q)]= G_R(m)[p_\mu
e_\nu(\phi) - p_\nu e_\mu(\phi)][k_\mu e_\nu(\gamma) - k_\nu
e_\mu(\gamma)].$
Consequently, the function\\[0.1cm]
$g_R(m)= - 2(pk)G_R(m) = - 2\omega (m) m_\phi G_R(m)$\\[0.1cm]
\begin{figure}\includegraphics[width=17pc]{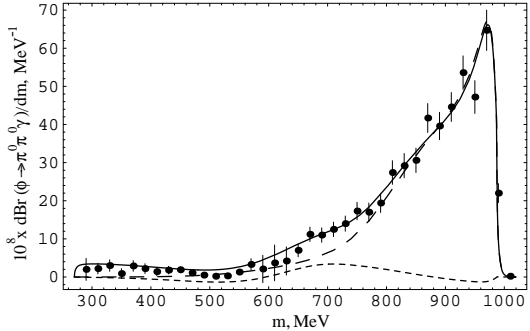}
\caption{\small The fit to the KLOE data for the $\pi^0\pi^0$ mss
spectrum in the $\phi\to\gamma\pi^0\pi^0$ decay.}\end{figure} is
proportional to the photon energy
$\omega(m)=$\\$(m_{\phi}^2-m^2)/2m_{\phi}$ (at least!) in the soft
photon region. Stopping the function $(\omega (m))^2$ at $\omega
(990\,\mbox{MeV})=29$ MeV with the help of the form-factor $1/\left
[1+(R\omega (m))^2\right ]$ requires $R\approx 100$ GeV$^{-1}$. It
seems to be incredible to explain such a huge radius in hadron
physics. Based on rather great $R\approx 10$ GeV$^{-1}$, one can
obtain an effective maximum of the mass spectrum only near 900 MeV.
To exemplify this trouble let us consider the contribution of the
isolated $R$ resonance: $g_R(m)=-2\omega (m) m_\phi G_R\left
(m_R\right)$. Let also the mass and the width of the $R$ resonance
equal 980 MeV and 60 MeV, then $S_R(920\,\mbox{MeV}):
S_R(950\,\mbox{MeV}):S_R(970\,\mbox{MeV}):S_R(980\,\mbox{MeV})
=3:2.7:1.8:1$.So stopping the $g_R(m)$ function is the crucial point
in understanding  the mechanism  of the production of $a_0(980)$ and
$f_0(980)$ resonances in the $\phi$ radiative decays. The
$K^+K^-$-loop model $\phi\to K^+K^-\to\gamma R$ solves this problem
in the elegant way: fine threshold phenomenon is discovered, see
Fig. 7. So, the mechanism of production of $a_0(980)$ and $f_0(980)$
mesons in the $\phi$ radiative decays is established
\begin{figure}\includegraphics[width=17pc]{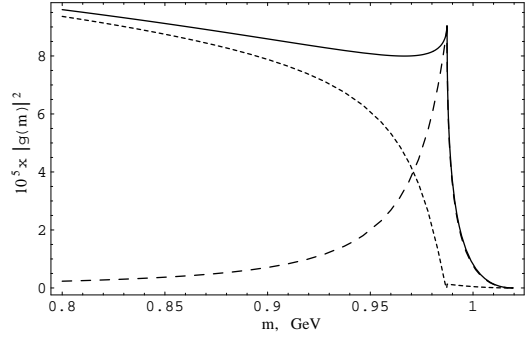}
\caption{\small The universal in the $K^+K^-$ loop model function
$|g(m)|^2=|g_R(m)/g_{RK^+K^-}|^2$ is shown by the solid curve. The
contribution of the immaginary (real) part is shown by dashed
(dotted) curve.}\end{figure} at a physical level of proof, see Refs.
\cite{A2,AG} for details. This production mechanism is the
four-quark transition what constrains the large $N_C$ expansion of
the $\phi\to\gamma a_0(980)$ and $\phi\to\gamma f_0(980)$ amplitudes
and gives the new strong (if not crucial) evidences in favor of the
four-quark nature of $a_0(980)$ and $f_0(980)$ mesons \cite{A2}.

\section{THE \boldmath{$J/\psi$} DECAYS ABOUT NATURE OF LIGHT SCALAR RESONANCES \cite{A1}}

{\bf \boldmath $a_0(980)$ in $J/\psi$ decays.} The following data is
of very interest for our purposes: $B(J/\psi\to a_0(980)\rho) <
4.4\cdot 10^{-4}$ and $B(J/\psi\to a_2(1320)\rho)= (109\pm 22)\cdot
10^{-4}$. The suppression $B(J/\psi\to a_0(980)\rho)/B(J/\psi\to
a_2(1320)\rho)<0.04\pm0.008$ seems strange, if one considers the
$a_2(1320)$ and $a_0(980)$ states as the tensor and scalar isovector
states from the same $P$-wave $q\bar q$ multiplet. While the
four-quark nature of the $a_0(980)$ meson with the symbolic quark
structure, similar (but, generally speaking, not identical) the
MIT-bag state, $a_0^+ = us\bar d\bar s$, $a_0^0=(us\bar u\bar
s-ds\bar d\bar s)/\sqrt{2}$, $a_0^- = ds\bar u\bar s$, is not
contrary to the suppression under discussion. So, the improvement of
the upper limit for $B(J/\psi\to a_0(980)\rho)$ and the search for
the $J/\psi\to a_0(980)\rho$ decays  are the urgent purposes in the
study of the $J/\psi$ decays!

Recall that twenty years ago the four-quark nature of $a_0(980)$ was
supported by suppression of $a_0(980)\to\gamma\gamma$ as was
predicted in our work based on the $q^2\bar q^2$ model,
$\Gamma(a_0(980)\to\gamma\gamma)\sim 0.27\,\mbox{keV}$. Experiment
gives $\Gamma (a_0\to\gamma\gamma)=(0.19\pm 0.07
^{+0.1}_{-0.07})/B(a_0\to\pi\eta)$ keV, Crystal Ball, and $\Gamma
(a_0\to\gamma\gamma)=(0.28\pm 0.04\pm 0.1)/B(a_0\to\pi\eta)$ keV,
JADE. When in the $q\bar q$ model it was anticipated
$\Gamma(a_0\to\gamma\gamma)=(1.5 - 5.9)\Gamma (a_2\to\gamma\gamma)=
(1.5 - 5.9)\cdot(1.04\pm 0.09)$ keV.

{\bf \boldmath $f_0(980)$ in $J/\psi$ decays.} The hypothesis that
the $f_0(980)$ meson is the lowest two-quark $P$ wave scalar state
with the quark structure $f_0(980)=(u\bar u+d\bar d)/\sqrt{2}$
contradicts the following facts. 1) The strong coupling with the
$K\bar K$-channel, $1<|g_{f_0K^+K^-}/g_{f_0\pi^+\pi^-}|^2<10$, for
the prediction $|g_{f_0K^+K^-}/g_{f_0\pi^+\pi^-}|^2=\lambda/4\simeq
1/8$. 2) The weak coupling with gluons, $B(J/\psi\to\gamma
f_0(980)\to\gamma\pi\pi)< 1.4\cdot10^{-5}$, opposite the expected
one $B(J/\psi\to\gamma f_0(980))\approx B(J/\psi\to\gamma
f_2(1270))/4=(3.45\pm 0.35)\cdot 10^{-4}$. 3) The weak coupling with
photons, predicted in our work for the $q^2\bar q^2$ model,
$\Gamma(f_0(980)\to\gamma\gamma)\sim 0.27\,\mbox{keV}$, and
supported by experiment, $\Gamma (f_0\to\gamma\gamma)=(0.31\pm
0.14\pm 0.09)$ keV, Crystal Ball, and $\Gamma
(f_0\to\gamma\gamma)=(0.24\pm 0.06\pm 0.15)$ keV, MARK II. When in
the $q\bar q$ model it was anticipated
$\Gamma(f_0\to\gamma\gamma)=(1.7 - 5.5)\Gamma (f_2\to\gamma\gamma)=
(1.7 - 5.5)\cdot(2.8\pm 0.4)$ keV. 4) As is the case with $a_0(980)$
the suppression $B(J/\psi\to f_0(980)\omega) /B(J/\psi\to
f_2(1270)\omega)= 0.033\pm 0.013$ looks strange in the model under
consideration. We should like to emphasize that from our point of
view the DM2 Collaboration did not observed the $J/\psi\to
f_0(980)\omega$ decay and should give a upper limit only. So, the
search for the $J/\psi\to f_0(980)\omega$ decay is the urgent
purpose in the study of the$J/\psi$ decays! The existence of the
$J/\psi\to f_0(980)\phi$ decay of greater intensity than the
$J/\psi\to f_0(980)\omega$ decay shuts down the $f_0(980)=(u\bar
u+d\bar d)/\sqrt{2}$ model. In the case under discussion the
$J/\psi\to f_0(980)\phi$ decay should be strongly suppressed in
comparison with the $J/\psi\to f_0(980)\omega$ decay by the OZI
rule.

Can one consider the $f_0(980)$ meson as the near $s\bar s$-state?
It is impossible without a gluon component. Really, it is
anticipated for the scalar $s\bar s$-state from the lowest P-wave
multiplet that $B(J/\psi\to\gamma f_0(980))\approx B(J/\psi\to\gamma
f_2^\prime(1525))/4=(1.175 ^{+0.175}_{-0.125})\cdot 10^{-4}$
opposite experiment $< 1.4\cdot 10^{-5}$, which requires properly
that the $f_0(980)$-meson to be the 8-th component of the $SU_f(3)$
oktet $f_0(980)=(u\bar u+d\bar d-2s\bar s)/\sqrt{6}$. But this
structure gives $B(J/\psi\to f_0(980)\phi)=(2\lambda\approx 1)\cdot
B(J/\psi\to f_0(980) \omega)$ which is on the verge of conflict with
experiment. Here $\lambda$ takes into account the strange sea
suppression. The $SU_f(3)$ oktet case  contradicts also the strong
coupling with the $K\bar K$ channel
$1<|g_{f_0K^+K^-}/g_{f_0\pi^+\pi^-}|^2<10$ for the prediction
$|g_{f_0K^+K^-}/g_{f_0\pi^+\pi^-}|^2=(\sqrt{\lambda}-2)^2/4\approx
0.4$. In addition, the mass degeneration $m_{f_0}\approx m_{a_0}$ is
coincidental in this case if to treat the $a_0$-meson as the
four-quark state or contradicts the two-quark hypothesis.

The introduction of a gluon component, $gg$, in the $f_0(980)$ meson
structure  allows the puzzle of weak coupling with two gluons and
with two photons but the strong coupling with the $K\bar K$ channel
to be resolved easily: $f_0=gg\sin\alpha\,+$\,$[(1/\sqrt{2})(u\bar
u+d\bar d)\sin\beta+s\bar s\cos\beta]\cos\alpha$,
$\tan\alpha=-O(\alpha_s)(\sqrt{2}\sin\beta +\cos\beta),$ where
$\sin^2\alpha\leq 0.08$ and $\cos^2\beta
> 0.8$. So, the $f_0(980)$ meson is  near to the $s\bar s$-state. It
gives $$0.1 <\frac{B(J/\psi\to f_0(980)\omega)}{B(J/\psi\to
f_0(980)\phi)}= \frac{1}{\lambda}\tan^2\beta<0.54\,.$$ As for the
experimental value, $B(J/\psi\to f_0(980)\omega)/B(J/\psi\to
f_0(980)\phi)=0.44\pm 0.2$, it needs refinement. Remind that in our
opinion the $J/\psi\to f_0(980)\omega$ was not observed!

The scenario with the $f_0(980)$ meson near to the $s\bar s$ state
and with the $a_0(980)$ meson as the two-quark state runs into
following difficulties. 1) It is impossible to explain the $f_0$ and
$a_0$-meson mass degeneration in a natural way. 2) It is predicted
$\Gamma(f_0\to\gamma\gamma)<0.13\cdot\Gamma(a_0\to\gamma\gamma)$,
that means that $f_0(980)$ could not be seen practically in the
$\gamma\gamma$ collision. 3) It is predicted $B(J/\psi\to
a_0(980)\rho)=(3/\lambda\approx 6)\cdot B(J/\psi\to f_0(980)\phi)$,
that has almost no chance from experimental point view. 4) The
$\lambda$ independent prediction $B(J/\psi\to
f_0(980)\phi)/B(J/\psi\to f_2'(1525)\phi)=B(J/\psi\to
a_0(980)\rho)/B(J/\psi\to a_2(1320)\rho)<0.04\pm 0.008$ is excluded
by the central figure in $B(J/\psi\to f_0(980)\phi) /B(J/\psi\to
f_2'(1525)\phi)= 0.4\pm0.23$. But, certainly, experimental error is
too large. Even twofold increase in accuracy of measurement of could
be crucial in the fate of the scenario under discussion.

{\bf Summary on \boldmath $J/\psi$ decays.} The prospects for the
model of the $f_0(980)$ meson as the almost pure $s\bar s$-state and
the $a_0(980)$-meson as the four-quark state with the coincidental
mass degeneration is rather poor especially as the OZI-superallowed
$\left (N_C\right )^0$ order mechanism $\phi = s\bar s\to\gamma
s\bar s = \gamma f_0(980)$ \footnote{Such a mechanism is similar to
the principal mechanism of the $\phi\to\gamma\eta '(958)$ decay:
$\phi = s\bar s\to\gamma s\bar s = \gamma\eta'(958)$.} cannot
explain the photon spectrum in $\phi\to\gamma
f_0(980)\to\gamma\pi^0\pi^0$ \cite{A2}, which requires the
domination of the $K^+K^-$ intermediate state in the $\phi\to\gamma
f_0(980)$ amplitude: $\phi\to K^+K^-\to\gamma f_0(980)$! The $\left
(N_C\right )^0$ order transition is bound to have a small weight in
the large $N_C$ expansion of the $\phi = s\bar s\to\gamma f_0(980)$
amplitude, because this term does not contain the $K^+K^-$
intermediate state, which emerges only in the next to leading term
of the $1/N_C$ order, i.e., in the OZI forbidden transition
\cite{A2}. While the four-quark model with the symbolic structure $
f_0(980) = (us\bar u\bar s+ds\bar d\bar s)/\sqrt{2}\cos\theta +
ud\bar u\bar d\sin\theta$, similar (but not identical) the MIT-bag
state, reasonably justifies all unusual features of the
$f_0(980)$-meson.

\section{NEW ROUND IN $\gamma\gamma\to\pi^+\pi^-$, THE BELLE DATA \cite{AS2}}

Recently, the Belle Collaboration succeeded in observing a clear
manifestation of the $f_0(980)$ resonance in the reaction
$\gamma\gamma\to\pi^+\pi^-$. This has been made possible owing to
the huge statistics and good energy resolution. Analyzing these data
we shown that the above $K^+K^-$ loop mechanism provides the
absolutely natural and reasonable scale of the $f_0(980)$ resonance
manifestation in the $\gamma\gamma\to\pi^+\pi^-$ reaction cross
sections (as well as in $\gamma\gamma\to\pi^0\pi^0$ and the
$a^0_0(980)$ in $\gamma\gamma\to\pi^0\eta$). For the $K^+K^-$ loop
mechanism, we obtained the $f_0(980)\to\gamma\gamma$ width averaged
by the resonance mass distribution in the $\pi\pi$ channel
$\langle\Gamma^{Born}_{f_0\to K^+K^-\to\gamma\gamma}
\rangle_{\pi\pi}\approx0.15$ keV. Furthermore, the $K^+K^-$ loop
mechanism of the $f_0(980)\to\gamma\gamma$ coupling is one of the
main factors responsible for the formation of the observed specific,
steplike, shape of the $f_0(980)$ resonance in the
$\gamma\gamma\to\pi^+\pi^-$ reaction cross section.

\end{document}